\documentclass[universe,communication,accept,pdflatex,moreauthors]{Definitions/mdpi}

\firstpage{1}
\pubvolume{1}
\issuenum{1}
\articlenumber{0}
\pubyear{2025}
\copyrightyear{2025}
\externaleditor{Firstname Lastname}
\datereceived{11 June 2025}
\daterevised{15 July 2025} 
\dateaccepted{17 July 2025}
\datepublished{ }
\usepackage{subcaption}
\usepackage[utf8]{inputenc}
\usepackage{graphicx}
\usepackage{booktabs}
\usepackage{geometry}
\usepackage{color}
\usepackage{bm}
\usepackage{multirow}
\usepackage{amsmath}
\makeatletter
\renewcommand{\linenumbers}{}
\makeatother

\hreflink{https://doi.org/} 

\Title{Gamma-Ray Bursts Calibrated by Using Artificial Neural Networks from the Pantheon+ Sample 
}

\TitleCitation{Gamma-Ray Bursts Calibrated by Using Artificial Neural Networks from the Pantheon+ Sample}

\Author{Zhen 
 Huang $^{1,\dagger }$\orcidA{},  Xin Luo $^{1,\dagger }$\orcidB{}, Bin Zhang $^{2}$\orcidC{}, Jianchao Feng $^{3}$, Puxun Wu $^{4}$\orcidE{}, Yu Liu $^{5}$\orcidF{} and Nan Liang $^{1,}$*\orcidG{}}

\AuthorNames{Zhen Huang, Xin Luo, Bin Zhang, Jianchao Feng, Puxun Wu,  Yu Liu and Nan Liang}

\AuthorCitation{Huang, 
 Z.; Luo, X.; Zhang, B.; Feng, J.; Wu, P.; Liu, Y.; Liang, N.}

\address{%
$^{1}$ \quad Guizhou Key Laboratory of Advanced Computing, Guizhou Normal University, Guiyang 550025, China; 
 zhen\_huang@gznu.edu.cn (Z.H.); xin\_luo@gznu.edu.cn (X.L.) 
 \\
$^{2}$ \quad School of Mathematical Sciences, Guizhou Normal University, Guiyang 550025, China;  binzhang@gznu.edu.cn 
\\
$^{3}$ \quad Department of Astronomy, Guizhou Normal University, Guiyang 550025, China;  fengjc@gznu.edu.cn\\
$^{4}$ \quad Department of Physics and 
 Synergistic Innovation Center for Quantum Effects and Applications, Hunan Normal University, Changsha 410081, China; pxwu@hunnu.edu.cn 
  \\
$^{5}$ \quad School of Physical Science and Technology, Southwest Jiaotong University, Chengdu 611756,  China; lyu@swjtu.edu.cn} 

\corres{Correspondence: liangn@bnu.edu.cn}
\firstnote{These authors contributed equally to this work.}

\abstract{
  In this paper, we calibrate the luminosity relation of gamma$-$ray bursts (GRBs) by employing artificial neural networks (ANNs) to analyze the Pantheon+ sample of type Ia supernovae (SNe Ia) in a manner independent of cosmological assumptions. The A219 GRB dataset is used to calibrate the Amati relation (\(E_{\rm p}\)-\(E_{\rm iso}\)) at low redshift with the ANN framework, facilitating the construction of the Hubble diagram at higher redshifts. Cosmological models are constrained with GRBs at high redshift and the latest observational Hubble data (OHD) via the Markov chain Monte Carlo numerical approach. For the Chevallier$-$Polarski$-$Linder (CPL) model within a flat universe, we obtain \(\Omega_{\rm m} = 0.321^{+0.078}_{-0.069}\), \(h = 0.654^{+0.053}_{-0.071}\), \(w_0 = -1.02^{+0.67}_{-0.50}\), and \(w_a = -0.98^{+0.58}_{-0.58}\) at the 1~$-$~\(\sigma\)~zconfidence level, which indicates a preference for dark energy with potential redshift evolution (\(w_a \neq 0\)). These findings using ANNs align closely with those derived from GRBs calibrated using Gaussian processes (GPs).
	}


\begin{document}
\section{Introduction}

Gamma$-$ray 
 bursts (GRBs) can serve as cosmic probes by leveraging luminosity relations
to explore the universe's expansion history 
at redshifts far beyond Type Ia supernovae (SNe Ia) \citep{Schaefer2003,Dai2004,Ghirlanda2004b,Firmani2005,Xu2005,Liang2006,Wang2006,Ghirlanda2006,Schaefer2007}.
To address the circularity problem, Ref.
~\cite{Liang2008} proposed a model$-$independent method to calibrate seven GRB luminosity relations utilizing SNe Ia at low redshifts.
In~\cite{Amati2019}, the authors used observational Hubble data (OHD) obtained via the cosmic chronometers (CC) method to calibrate the Amati relation, which connects the spectral peak energy to the isotropic equivalent radiated energy  
\citep{Amati2002}.  
Therefore, GRB data  can be used to constrain cosmological models at high redshift by using the standard Hubble diagram method \citep{Capozziello2008,Capozziello2009,Wei2009,Wei2010,Liang2010,Liang2011,Demianski2017a,Demianski2017b,Montiel2021,Luongo2023,Wang2016}.
Other local data  have also been used  to calibrate GRBs, e.g., mock data of gravitational waves (GWs) \citep{Wang2019}, quasars \citep{Dai2021,Purohit2024}, and angular diameter distances of galaxy clusters \citep{Gowri2022}.
In addition,Ref. \cite{Amati2008} proposed the simultaneous fitting method, which constrains the coefficients of the relationship and the parameters of the cosmological model simultaneously to alleviate the circularity problem. 
It has been found that the Amati relation parameters are almost identical in all cosmological models by the simultaneous fitting method with a dataset of 118 GRBs (the A118 sample) from the total 220 GRBs (the A220 sample) \citep{Khadka2020,Khadka2021,Cao2022a,Cao2022b,Cao2022}. 
Recent works on the luminosity relations of GRBs and their applications for cosmology can be found in \cite{Hu2021,Liu2022a,Liu2022b,Tian2023,Li2024,Han2024,Favale2024,Colgain2024,Paliathanasis2025}; see \cite{Bargiacchi2025,Deng2025} for reviews.

Similar to the interpolation method used in \cite{Liang2008} and the B\'ezier parametric used in~\cite{Amati2019}, GRBs can be calibrated from local data using  iterative procedures \citep{LiangZhang2008}, polynomial fitting \citep{Kodama2008}, local regression \citep{Cardone2009,Demianski2017a}, cosmography methods \citep{Capozziello2010,Gao2012}, the Pad\'e approximation method \citep{Liu2015}, or a two$-$step method \citep{Izzo2015,Muccino2021}.
Recently, Gaussian processes (GPs) 
\citep{Seikel2012a} have been used in GRB cosmological studies \citep{Liang2022,Li2023,Mu2023,WL2024,Wang2024,Nong2024,Xie2025,Seikel2012b}. 
However, in GP analysis it is typically assumed that the errors in observational data follow a Gaussian distribution~\citep{Seikel2012a}, which may pose a substantial limitation when reconstructing functions from data;
furthermore, the results can be affected by the choice of the kernel functions, with many different available kernel functions \citep{Wei2017,Zhou2019}.
The application of machine learning (ML) techniques has revolutionized data analysis in cosmology, offering robust tools for reconstructing complex astrophysical relationships and constraining cosmological parameters. 
In~\cite{Luongo2021b}, the authors explored three ML treatments (linear regression, neural network, random forest) to alleviate the circularity problem with the Amati relation.
In~\cite{Bengaly2023}, machine learning algorithms were deployed to measure $H_0$ through regression analysis, 
finding that Support Vector Machine (SVM) exhibited the best performance in terms of bias$-$variance tradeoff in most cases, showing itself to be a competitive cross$-$check to GP.
In~\cite{Zhang2025}, the authors utilized high$-$performance KNN (K-Nearest Neighbors) and RF (Random Forest) machine learning algorithms based on the Pantheon+ dataset \citep{Scolnic2022} and the A219 sample \citep{Khadka2021,Liang2022} 
to calibrate the Amati relation in a model$-$independent manner and construct the Hubble diagram.
By combining high$-$redshift data and observational Hubble data to constrain cosmological models, their results are consistent with those calibrated by Gaussian processes, providing a new pathway for precise cosmological studies.

Moreover, ANNs excel in modeling nonlinear correlations without requiring predefined functional forms, which can be used for cosmology at high redshifts 
\citep{Escamilla-Rivera2020,Wang2020,Dialektopoulos2022,Gomez-Vargas2023,Zhang2024,Chen2024,Niu2025,Di2025}.
In the context of GRBs, ANN$-$based approaches can provide a powerful framework for calibrating empirical luminosity relations. 
In~\cite{Shah2024}, the  authors used a novel deep learning framework to reconstruct the cosmic distance ladder with a GRB sample, while~\cite{Mukherjee2024} employed neural networks to calibrate the Dainotti relations
 \endnote{The 2D Dainotti relation  \citep{Dainotti2008} is the correlation between the plateau luminosity and its end time in X$-$ray afterglows; the 3D Dainotti relation  \citep{Dainotti2016} is the correlation incorporating the peak prompt luminosity with the plateau end time and luminosity in the rest frame, achieving a small intrinsic scatter.} from the Pantheon+ sample of SNe Ia. 
In~\cite{HuangLiang2025}, an ANN framework was employed based on observational Hubble data (OHD) from cosmic chronometers, reconstructing $H(z)$ in a model$-$independent way for relation calibration. Considering the physical correlations in the data, the authors introduced the covariance matrix and Kullback$-$Leibler (KL) divergence into the loss function, then used the A219 \citep{Liang2022,Khadka2021}  and J220 samples \citep{Cao2024,Jia2022}  to select the optimal ANN model for calibrating the Amati relation.

Recent advances have incorporated machine learning approaches such as ANNs in combination with Bayesian neural networks (BNNs) to enhance the precision and reliability of these calibrations while quantifying uncertainties in a cosmology$-$independent \mbox{manner~\citep{Escamilla-Rivera2022,Tang2021,Tang2022}.}
In this study, we employ a hybrid ANN+BNN framework to calibrate the Amati relation. We utilize the Pantheon+ sample of SNe Ia 
to construct a high$-$redshift Hubble diagram and derive constraints on cosmological parameters in a flat universe, offering insights into the evolution of dark energy.

\section{Reconstructing the Apparent Magnitude Redshift Relation from~Pantheon+ Data}

\begin{figure}[H]
	\hspace{-4pt}\includegraphics[width=0.95\textwidth]{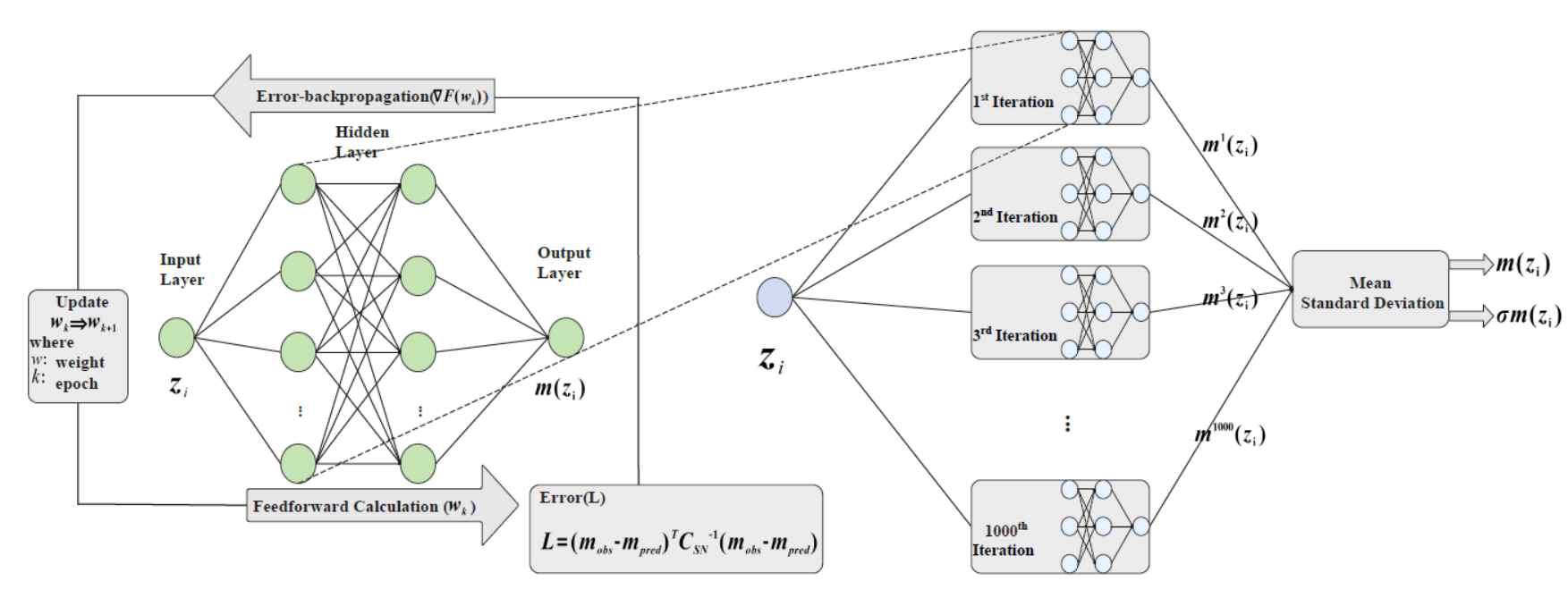}
	\caption{Architecture 
 of the ANN+BNN framework for fitting Pantheon+ SNe Ia apparent magnitudes $m(z_i)$. The left panel depicts the ANN structure, which maps the redshift $z_i$ to $m(z_i)$. The right panel shows the BNN simulation, where the ANN with dropout is executed over 1000 iterations for a given $z_i$. The mean of these predictions provides $m(z_i)$, while the standard deviation yields the uncertainty $\sigma_{m(z_i)}$.}
	\label{model}
\end{figure}

We explore the apparent magnitude$-$redshift relation using a hybrid model that integrates artificial (ANN) and Bayesian (BNN) neural networks. The ANN component models the relationship between redshift and apparent magnitude, while the BNN quantifies predictive uncertainties, accounting for the covariance errors inherent in the SNe Ia observations.
Through backpropagation \citep{Rumelhart1986}, the network iteratively refines its weights to minimize the loss function,~\endnote{
	We incorporate the Pantheon+ covariance matrix $C_{\rm SN}$ into the loss function:
	$
	\mathcal{L}_{\rm SN} = \Delta m(z_i)^T C_{\rm SN}^{-1} \Delta m(z_i),
	$
	where $\Delta m(z_i) = m(z_i)_{\rm pred} - m(z_i)_{\rm obs}$ represents the difference between predicted and observed magnitudes.} ensuring robust predictions.
While ANNs excel in capturing complex patterns, they lack inherent uncertainty quantification. To address this, we employ a BNN approach, utilizing dropout to approximate Bayesian inference \citep{Gal2016a,Gal2016b}. The ANN with dropout is run for 1000 iterations, producing a distribution of predicted magnitudes $m^j(z_i)$ 
for each redshift. The mean of these predictions serves as the final $m(z_i)$, with the standard deviation providing the uncertainty $\sigma_{m(z_i)}$.
The ANN processes redshift inputs $z_i$ to generate corresponding apparent magnitudes $m(z_i)$, as illustrated in Figure \ref{model}.

\vspace{-6pt}

Effective model performance depends on well$-$tuned hyperparameters, including the batch size, number and structure of hidden layers, activation function, and dropout rate, which together mitigate overfitting and enhance generalization ability \citep{Srivastava2014}.
We performed a grid search across 360 combinations, as summarized in Table \ref{Mul ML}, selecting the configuration with the lowest loss. This configuration consisted of a batch size of 16, one hidden layer with 512 units, a Tanh activation function, and a dropout rate of 0.1.
\begin{table}[H]

	\caption{Candidate 
 hyperparameters for the ANN+BNN framework used to model Pantheon+ SNe Ia apparent magnitudes $m(z_i)$. Optimal values determined via grid search are highlighted in bold. \label{Mul ML}}
	\begin{tabularx}{\textwidth}{cCC}
		\toprule
		\textbf{Hyperparameter} & \multicolumn{2}{c}{\textbf{Candidate} \textbf{Values}} \\
		\midrule
		Batch size  & \multicolumn{2}{c}{\textbf{16}, 32, 64} \\
		\midrule
		\multirow{4}[8]{*}{Hidden layers} & \multicolumn{1}{c}{Layers} & Units \\
		\cmidrule{2-3}          & \multicolumn{1}{c}{\textbf{1}} & 64, 128, 256, \textbf{512} \\
		\cmidrule{2-3}          & \multicolumn{1}{c}{2} & (64, 32), (128, 64) \\
		\cmidrule{2-3}          & \multicolumn{1}{c}{3} & (256, 128, 64), (512, 256, 128) \\
		\midrule
		Activate function & \multicolumn{2}{c}{ReLU, Sigmoid, \textbf{Tanh}} \\
		\midrule
		Dropout rate & \multicolumn{2}{c}{\textbf{0.1}, 0.2, 0.3, 0.4, 0.5} \\
		\bottomrule
	\end{tabularx}
\end{table}

\begin{figure}[H]
	\hspace{-3pt}\includegraphics[width=0.7\textwidth]{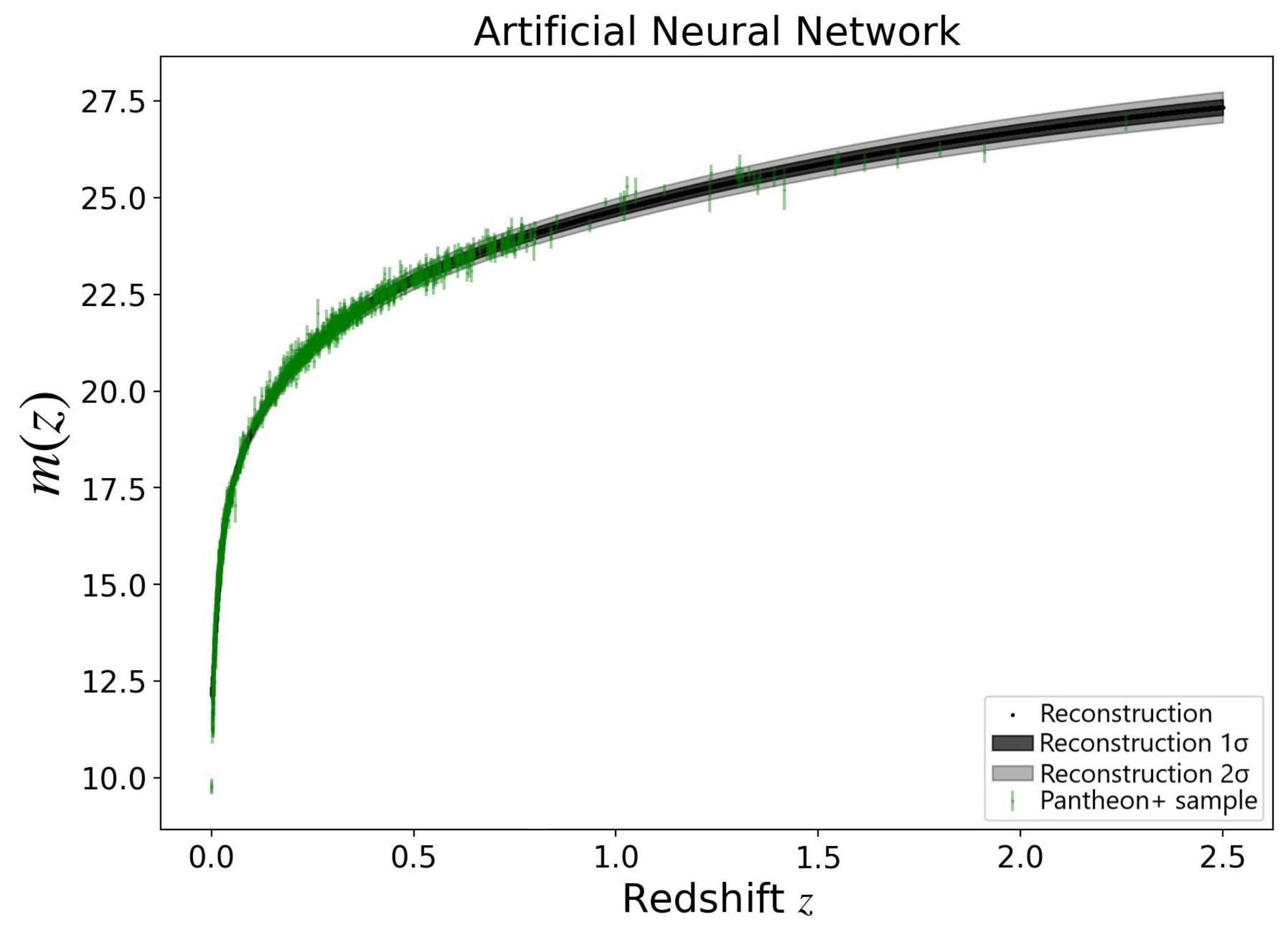}
	\caption{Reconstruction of the relation between the apparent magnitude and the redshift from the Pantheon+ dataset using the proposed ANN+BNN. Green dots indicate Pantheon+ data points with 1$\sigma$ error bars. The black line represents the reconstructed central value, with shaded regions denoting 1$\sigma$ and 2$\sigma$ uncertainties. 
	}
	\label{recon}
\end{figure}

The ANN+BNN reconstruction for fitting Pantheon+ SNe Ia apparent magnitudes $m(z_i)$ are shown in Figure \ref{recon}.
In this work, we utilize the A219  sample \citep{Liang2022},\endnote{The A219  sample is refined from the A220 sample \citep{Khadka2021} by removing the GRB051109A. 
} which is divided into a low$-$redshift subset ($z < 1.4$, 79 GRBs) for calibration and a high-redshift subset ($z \geq 1.4$, 182 GRBs) for cosmological analysis.

\section{Calibration of Amati Relation}
The Amati relation linking the spectral peak energy ($E_{\rm p}$) to the isotropic equivalent radiated energy ($E_{\rm iso}$) is formulated as follows:
\begin{equation}
	y = a + bx
\end{equation}
where $y = \log_{10} \frac{E_{\rm iso}}{1 \, \rm erg}$, $x = \log_{10} \frac{E_{\rm p}}{300 \, \rm keV}$, and $a$ and $b$ are free parameters. These quantities are defined as
$
E_{\rm iso} = 4\pi d_L^2(z) S_{\rm bolo} (1+z)^{-1}, \quad E_{\rm p} = E_{\rm p}^{\rm obs} (1+z),
$
with $d_L(z)$ as the luminosity distance, $S_{\rm bolo}$ as the bolometric fluence, and $E_{\rm p}^{\rm obs}$ as the observed spectral peak energy. The apparent magnitude $m$ can be used to calibrate the relations between X$-$ray luminosity ($L_X$) and ultraviolet luminosity ($L_{UV}$) of quasars without assuming a prior absolute magnitude of SNe Ia\endnote{The distance module of SN Ia is related to the luminosity distance and the absolute magnitude ($M$); the value of $M$ cannot be directly obtained using only the SN Ia sample, and as such $M$ is treated as a free parameter. } by introducing a new coefficient $\beta'$ as a free parameter \cite{Zhang2024b}. Furthermore, the Amati relation can be reformulated in terms of apparent magnitude $m$ and a new coefficient $a'$ as a free parameter \citep{Zhang2025}:
$
y' = a' + bx,
$
where $y' = \log_{10} \left[ (1+z)^{-1} \left( \frac{S_{\rm bolo}}{1 \, \rm erg \, cm^{-2}} \right) \right] + \frac{2}{5}m$ and $a' = a + 2 \left( \frac{M}{5} - 1 \right) - \log_{10} [4\pi (\rm pc/cm)^2]$. This permits calibration directly from observed GRB data using the apparent magnitude of SN Ia reconstructed at a redshift of GRBs at the same redshift.

We implement Markov chain Monte Carlo (MCMC) fitting via the emcee 
 package~\citep{ForemanMackey2013}.
Parameters are fitted using the likelihood method of \citep{Reichart2001},\endnote{Likelihood method of \citep{Reichart2001}: $\mathcal{L}_{\rm R} \propto \prod_{i=1}^{N_1} \frac{\sqrt{1+b^2}}{\sigma} \times \exp \left[ -\frac{[y'_i - y'(x_i, z_i; a', b)]^2}{2\sigma^2} \right],$
	where $\sigma = \sqrt{\sigma_{\rm int}^2 + \sigma_{y',i}^2 + b^2 \sigma_{x,i}^2}$ and the intrinsic scatter is $\sigma_{\rm int} = \sqrt{\sigma_{y',\rm int}^2 + b^2 \sigma_{x,\rm int}^2}$.} which can avoid bias in variable selection \citep{Amati2013,LZL2023}.
We also used the GP method for comparison, which was done via the GaPP package with a squared exponential covariance function \citep{Seikel2012a}.
Results for the low-redshift A219 sample ($z < 1.4$) are presented in 
Table \ref{Amati result}. 
We find that the ANN+BNN results align with those obtained by GP with the A219 sample ($z < 0.8$)  \citep{Mu2023} at 1$\sigma$ uncertainties, confirming the efficacy of machine learning approaches. 
We also find that the quality of the uncertainty estimates produced here are not well$-$calibrated; the value of the intrinsic scatter obtained by our method in this case is slightly larger than other values obtained in the literature. It should be noted that the
 KL divergence,  which considers the physical meaning represented by the reconstruction data with their uncertainties and the covariance matrix of data, should be  introduced into the loss function in order to correct potential miscalibrations \cite{HuangLiang2025}.
 Actually, given the critical role of uncertainty quantification in astrophysical inference, ensuring that ANNs produce reliable calibrated uncertainties is essential for their robust application in cosmology.


\begin{table}[H]

		\caption{Best$-$fitting parameters ($a'$, $b$, $\sigma_{\rm int}$) for the Amati relation in the A219 GRB sample at $z < 1.4$ (79 GRBs) by ANN and GaPP methods along with the likelihood \citep{Reichart2001}. \label{Amati result}}
		\begin{tabularx}{\textwidth}{CCCCCC}
			\toprule
		 \textbf{Methods} & \textbf{Datasets} & \boldmath{$a'$} & \boldmath{$b$} & \boldmath{$\sigma_{\rm int}$} \\
			\midrule
			ANN & 79 GRBs ($z < 1.4$) & $4.89^{+0.05}_{-0.05}$ & $1.99^{+0.12}_{-0.15}$ & $0.55$ \\
			GaPP  & 79 GRBs ($z < 1.4$) & $4.88^{+0.07}_{-0.07}$ & $2.25^{+0.16}_{-0.21}$ & $0.71$ \\  
			\bottomrule
		\end{tabularx}

\end{table}

\section{The GRB Hubble Diagram and Constraints on DE Models}

\begin{figure}[H]
	\includegraphics[width=0.98\textwidth]{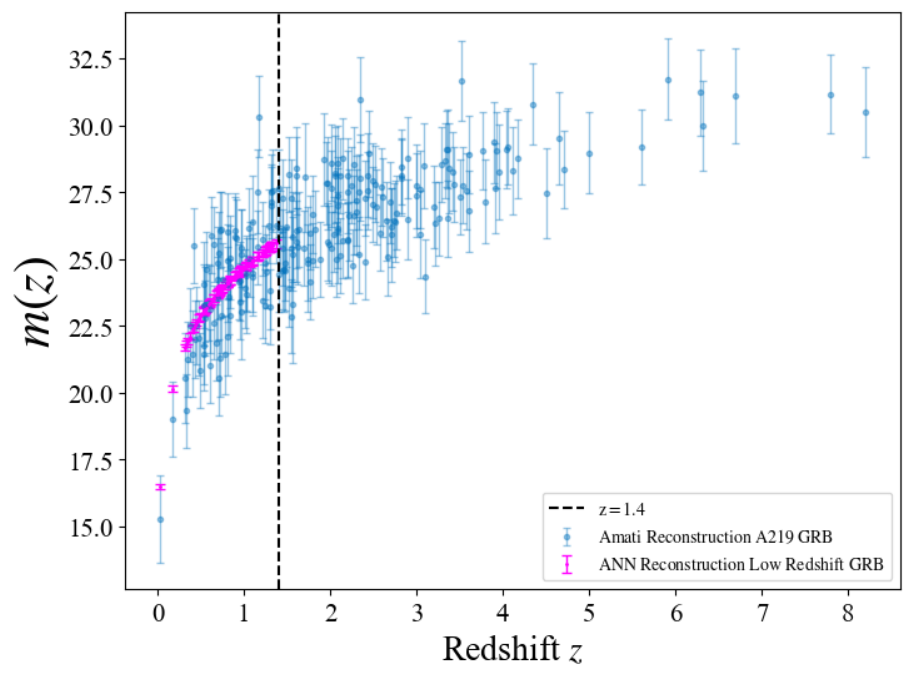}
	\caption{GRB 
 Hubble diagram for the A219 dataset. Purple points denote GRBs at $z < 1.4$ derived from Pantheon+ using the proposed ANN+BNN. 
	Blue points denote GRBs with the Amati relation calibrated using the likelihood method \citep{Reichart2001}, including low$-$redshift ($z < 1.4$) and high$-$redshift ($z \geq 1.4$) GRBs. 
}
	\label{GRB_Hubble}
\end{figure}

Assuming that the low$-$redshift calibration extends to higher redshifts, we construct the GRB Hubble diagram for $z \geq 1.4$. While the redshift dependence of GRB relations remains debated \citep{Lin2016,Wang2017,Demianski2017a,Demianski2021,Dai2021,Tang2021,Khadka2021,Liu2022a,Liu2022b,Jia2022,Kumar2023}, we apply the calibrated Amati relation, noting that evolutionary effects warrant further scrutiny.
By extrapolating the calibrated from the low$-$redshift GRBs to the high$-$redshift results using the ANN+BNN,  we can obtain GRB Hubble diagram.  The diagram combines low$-$redshift ($z < 1.4$) and high$-$redshift ($z \geq 1.4$) GRBs is shown in Figure \ref{GRB_Hubble}.\endnote{The uncertainty in the apparent magnitude is calculated as follows:
	$
	\sigma^2_m = \left( \frac{5}{2} \sigma_{y'}(a', b, x, \sigma_{\rm int}) \right)^2 + \left( \frac{5}{2 \ln 10} \frac{\sigma_{S_{\rm bolo}}}{S_{\rm bolo}} \right)^2,
	$
	where:
	$
	\sigma^2_{y'}(a', b, \sigma_{\rm int}, x) = \sigma^2_{\rm int} + \left( \frac{b}{\ln 10} \frac{\sigma_{E_{\rm p}}}{E_{\rm p}} \right)^2 + \sigma^2_{y'}(a', b)
	$
	and $\sigma^2_{y'}(a', b) = \left( \frac{\partial y'}{\partial a'} \right)^2 \sigma^2_{a'} + \left( \frac{\partial y'}{\partial b} \right)^2 \sigma^2_b + 2 \left( \frac{\partial y'}{\partial a'} \right) \left( \frac{\partial y'}{\partial b} \right) C^{-1}_{a'b}$, with $(C^{-1})_{a'b} = \frac{\partial^2 L}{\partial a' \partial b}$.}
To investigate dark energy (DE) properties, we leverage the high$-$redshift GRB Hubble diagram from the A219 sample 
to constrain the cosmological parameters in various DE models. 
We consider two flat models: the $\Lambda$CDM model with a constant equation of state (EoS) $w = -1$, and the Chevallier$-$Polarski$-$Linder (CPL) model \citep{CP2001,Linder2003} with a redshift$-$dependent EoS: $w(z) = w_0 + w_a z / (1 + z)$.\endnote{The luminosity distance in a flat universe is expressed as
	$
	d_{L;\rm th} = \frac{c (1 + z)}{H_0} \int_0^z \frac{dz'}{E(z')},
	$
	where
	$E(z) = [\Omega_{\rm m} (1 + z)^3 + \Omega_{\rm DE} (1 + z)^{3(1 + w_0 + w_a)} e^{-\frac{3 w_a z}{1 + z}}]^{1/2}$, 
	$\Omega_{\rm m}$ and $\Omega_{\rm DE}$ are respectively the matter and DE density parameters, with $\Omega_{\rm m} + \Omega_{\rm DE} = 1$ for flat geometry. For the $\Lambda$CDM model, $w_0=-1$ and $w_a=0$.}

\vspace{-3pt}

We fit the cosmological parameters by minimizing the $\chi^2$ statistic, incorporating the GRB covariance matrix $C_{\rm GRB}$. The $\chi^2$ function for GRBs is defined as follows: 
\begin{eqnarray}\label{chiGRB}
	\chi^2_{\rm GRB} = \Delta m_{\rm GRB}^T C_{\rm GRB}^{-1} \Delta m_{\rm GRB}
\end{eqnarray}
where $\Delta m_{\rm GRB} = m_{\rm GRB} - m_{\rm th}(P)$ is the residual vector between observed apparent magnitudes $m_{\rm GRB}$ and theoretical magnitudes $m_{\rm th}(P)$, and is calculated for cosmological parameters $P$. The theoretical magnitude is provided by
\begin{eqnarray}\label{mu}
	m_{\rm th}(P) = 5 \log_{10} \frac{d_L(P)}{\rm Mpc} + 25 + M = 5 \log_{10} D_L(P) - \mu_0 + M,
\end{eqnarray}
where $D_L(P) = d_L(P) H_0$ is the unanchored luminosity distance, $\mu_0 = 5 \log_{10} h + 42.38$, $h = H_0 / (100 \, \rm km/s/Mpc)$, and $H_0$ is the Hubble constant.

To enhance constraints, we incorporate 32 OHD measurements, including 31 Hubble parameter data points at $0.07 < z < 1.965$ \citep{Stern2010,Moresco2012,Moresco2015,Moresco2016,Zhang2014,Ratsimbazafy2017} and one additional point at \mbox{$z = 0.80$~\citep{Jiao2023}}.\endnote{An alternative OHD point at $z = 0.75$ is available \citep{Borghi2022}, but due to unclear covariance with \citep{Jiao2023}, we use only the latter, which accounts for a $1/\sqrt{2}$ fraction of systematic uncertainty. 
} The OHD $\chi^2$ is
\begin{equation}\label{eq:chi_Hz}
	\chi^2_{\rm OHD} = \Delta \hat{H}^T \mathbf{C}_H^{-1} \Delta \hat{H} + \chi^2_{\rm uncor},
\end{equation}
where $\Delta \hat{H} = H_{\rm th}(z; p) - H_{\rm obs}(z)$ for 15 correlated measurements \citep{Moresco2012,Moresco2015,Moresco2016}, $C_H^{-1}$ is the inverse covariance matrix \citep{Moresco2020}, and $\chi^2_{\rm uncor} = \sum_{i=1}^{17} [H_{\rm th}(z_i; p) - H_{\rm obs}(z_i)]^2 / \sigma_{H,i}^2$ for 17 uncorrelated measurements. The theoretical Hubble parameter is $H_{\rm th}(z; p) = H_0 \sqrt{\Omega_{\rm m} (1 + z)^3 + \Omega_{\rm DE} X(z)}$. The total $\chi^2$ is
\begin{equation}
	\chi^2_{\rm total} = \chi^2_{\rm GRB} + \chi^2_{\rm OHD}.
\end{equation}



\begin{figure}[H]
	\includegraphics[width=0.5\textwidth]{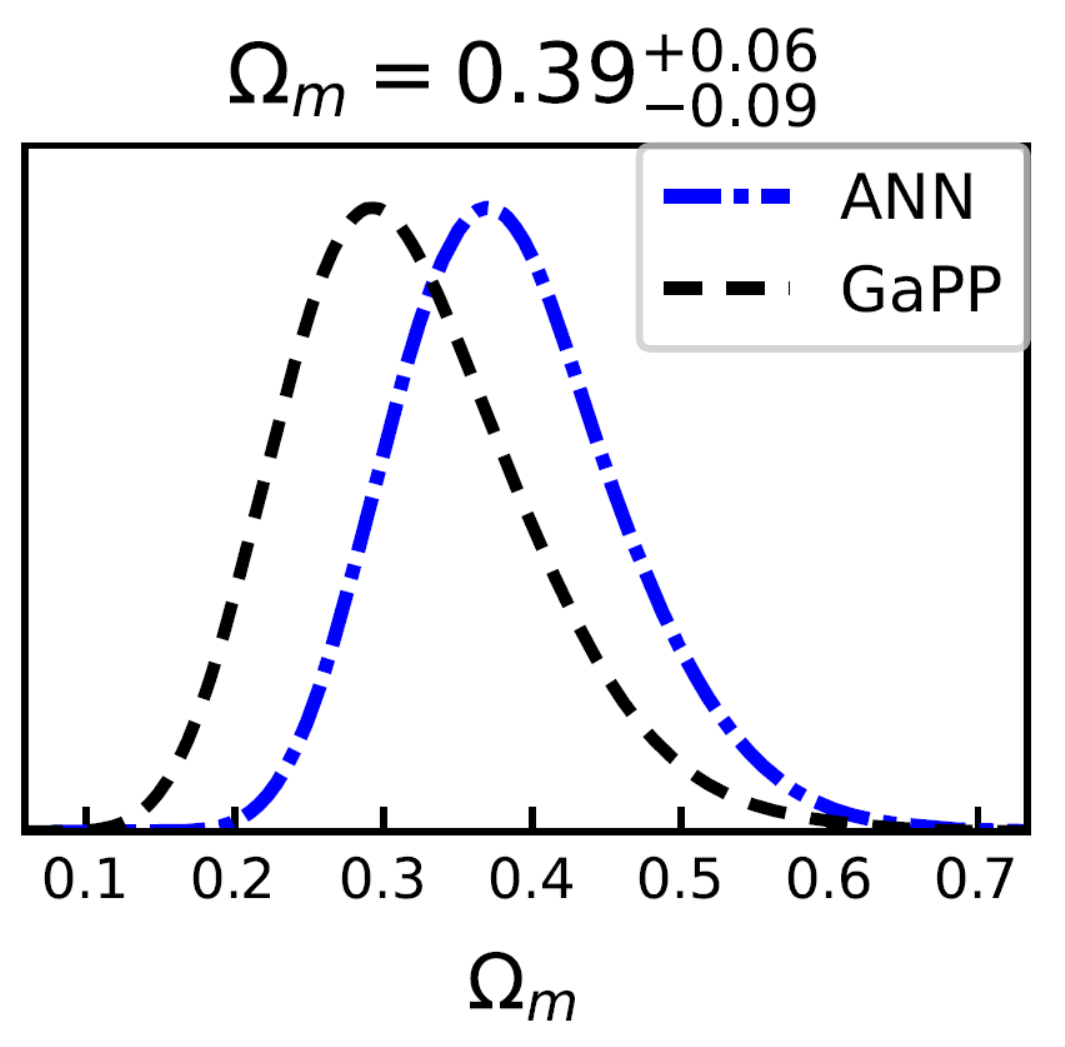}
	\caption{Constraints on $\Omega_{\rm m}$ for the flat $\Lambda$CDM model using 140 GRBs ($z > 1.4$) by ANN and GaPP methods, with $H_0$ fixed at 70 km/s/Mpc. }
	\label{constrain}
\end{figure}

\begin{figure}[H]
	\includegraphics[width=0.5\textwidth]{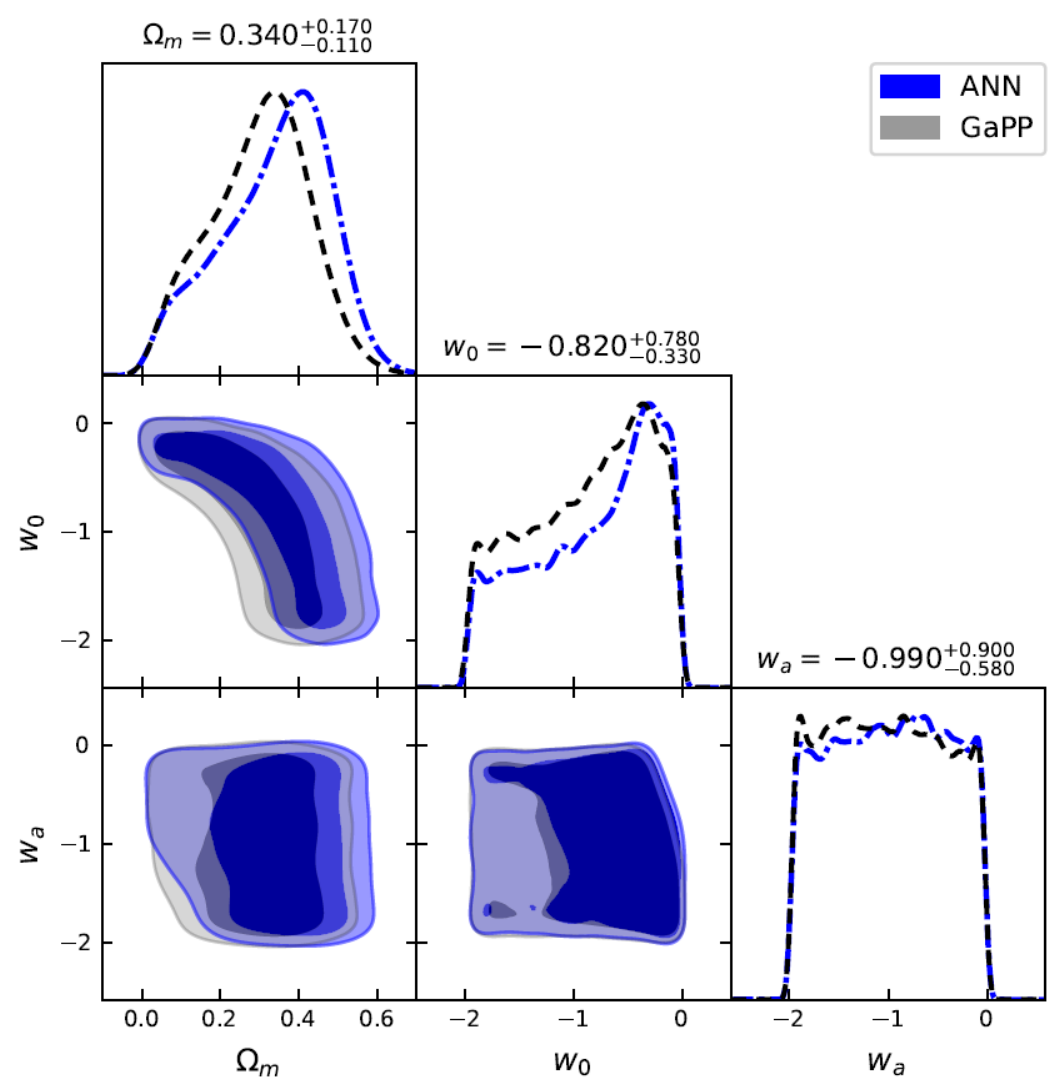}
	\caption{Constraints on $\Omega_{m}$, $w_0$, and $w_a$ for the flat CPL model using 140 GRBs ($z > 1.4$) by ANN and GaPP methods, with $H_0$ fixed at 70 km/s/Mpc. }
	\label{constrain_CPL}
\end{figure}

\begin{figure}[H]
	\includegraphics[width=0.5\textwidth]{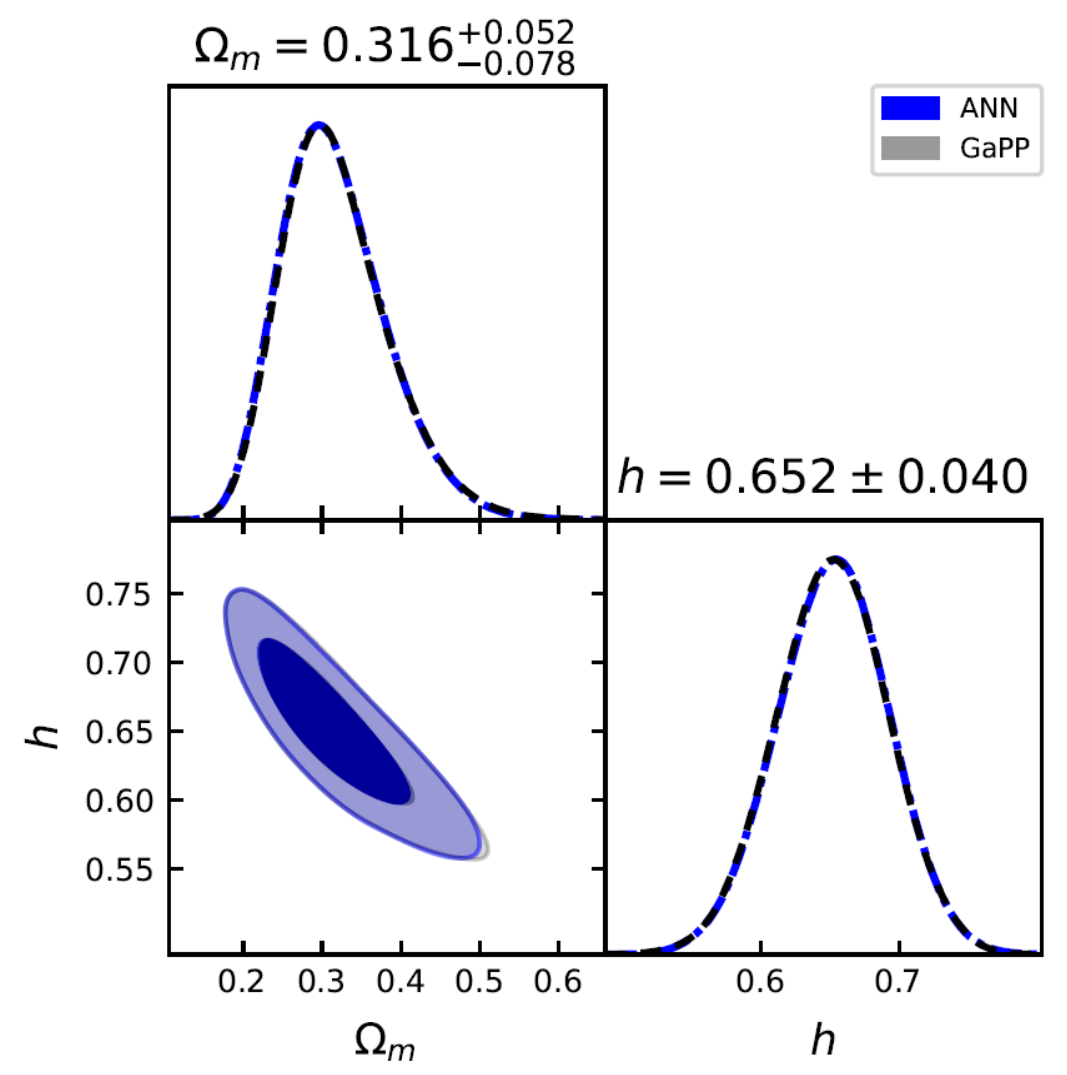}
	\caption{Joint constraints on $\Omega_{m}$ and $h$ for the flat $\Lambda$CDM model using 140 GRBs ($z > 1.4$) + 32 OHD by the ANN and GaPP methods. 
	}
	\label{Hubble_LambdaCDM}
\end{figure}

\begin{figure}[H]
	\includegraphics[width=0.5\textwidth]{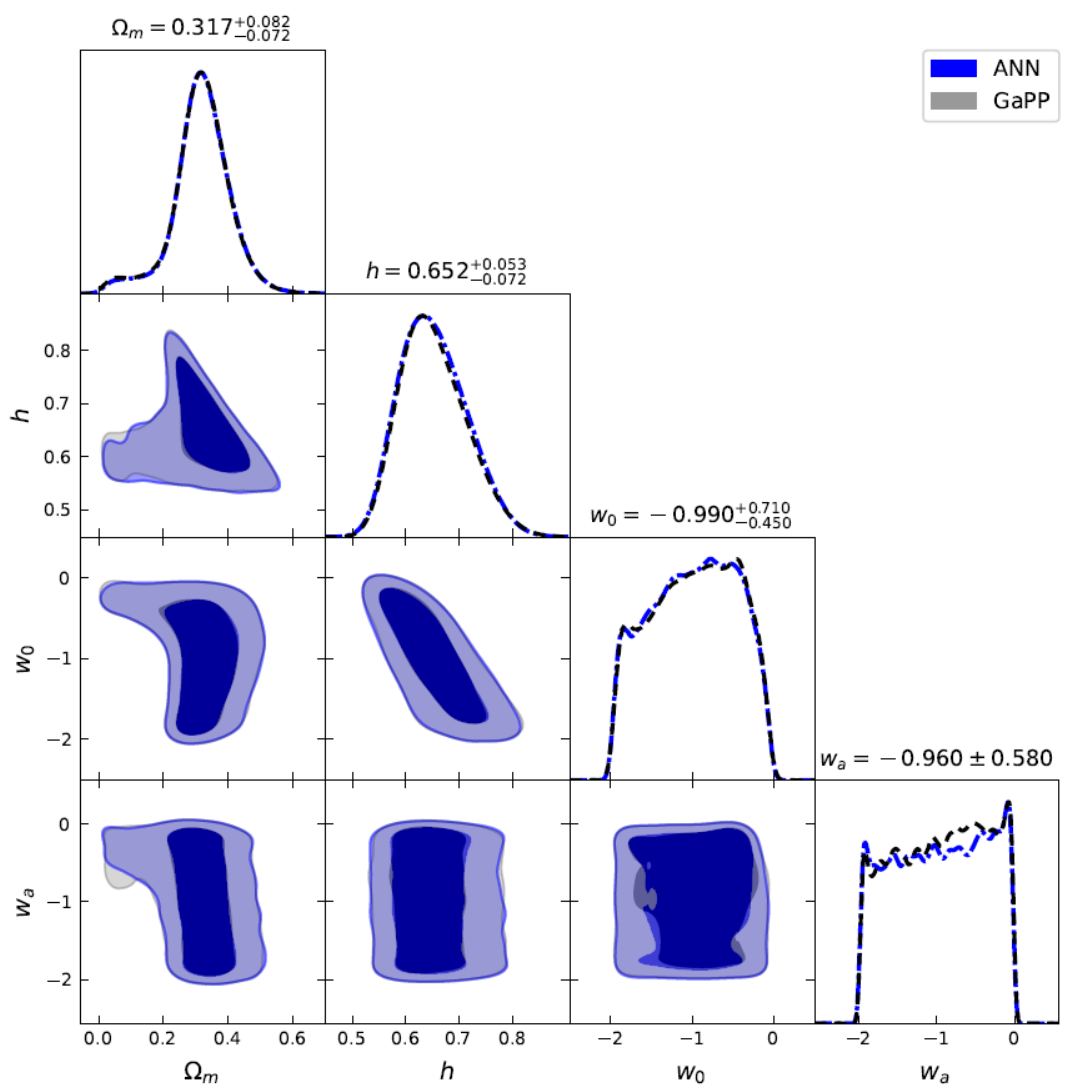}
	\caption{Joint constraints on $\Omega_{m}$, $h$, $w_0$, and $w_a$ for the flat CPL model using 140 GRBs ($z > 1.4$) + 32~OHD by the ANN and GaPP methods. }
	\label{Hubble_CPL}
\end{figure}

We perform MCMC fitting to constrain the DE models. Constraints using 140 GRBs at $z > 1.4$ are shown in Figures \ref{constrain} (for the $\Lambda$CDM model) and \ref{constrain_CPL} (for the CPL model); 
and joint constraints using 140 GRBs at $z > 1.4$  and 32 OHD are presented in Figures \ref{Hubble_LambdaCDM} (for $\Lambda$CDM model) 
and \ref{Hubble_CPL} (for CPL model), which are summarized in Table \ref{Joint constrain results}.
We find that the inclusion of OHD in the joint constraints can tighten the constraints significantly. 
The results for the CPL model at the 1$\sigma$ confidence level favor a possible DE evolution ($w_a\neq0$). 
We find that the ANN+BNN results are consistent with those obtained by GaPP, while showing slight differences. 
Compared to the fitting results from  CMB data based on the $\Lambda$CDM model at  very high redshift (\mbox{$H_0$ = 67.36 km s$^{-1}$ ${{\rm Mpc}}^{-1}$,} $\Omega_m$ = 0.315)~\citep{Planck2020} and SNe Ia at very low redshift (\mbox{$H_0$ = 74.3 km s$^{-1}$ ${{\rm Mpc}}^{-1}$,} $\Omega_m$ = 0.298)~\citep{Scolnic2022}, we find that the $H_0$  value with GRBs at \mbox{$1.4\le z\le8.2$} and OHD at $z\le1.975$  favors the one from the Planck observations and that the $\Omega_{m}$ value of our results for the flat $\Lambda$CDM model is consistent with the CMB observations at the 1$\sigma$ confidence level.
\vspace{-6pt}

\vspace{-6pt}
\vspace{-9pt}

\vspace{-9pt}
\begin{table}[H]
	\caption{Constraints on cosmological parameters for flat $\Lambda$CDM and CPL models by the ANN and GaPP methods with 
		GRBs ($z > 1.4$) only and GRBs ($z > 1.4$) + OHD. \label{Joint constrain results}}
	\footnotesize
\begin{adjustwidth}{-\extralength}{0cm}
\begin{tabularx}{\fulllength}{CClCCCCCCCC}
		\toprule
		\textbf{Models} & \textbf{Method} & \textbf{Data} \textbf{Sets} & \boldmath{$\Omega_{m}$} & \boldmath{$h$} & \boldmath{$w_0$} & \boldmath{$w_a$} & \boldmath{$-2\ln \mathcal{L}$} & \boldmath{$\Delta \rm AIC$} & \boldmath{$\Delta \rm BIC$ }\\
		\midrule
		\multirow{4}{*}{$\Lambda$CDM}
		& ANN & 140 GRBs & $0.385^{+0.062}_{-0.087}$ & - & - & - & 53.059 & - & - \\
		& GaPP & 140 GRBs & $0.318^{+0.064}_{-0.096}$ & - & - & - & 40.402 & - & - \\
		& ANN & 140 GRBs + 32 OHD & $0.316^{+0.052}_{-0.078}$ & $0.652^{+0.040}_{-0.040}$ & - & - & 78.576 & - & - \\
		& GaPP & 140 GRBs + 32 OHD & $0.318^{+0.051}_{-0.078}$ & $0.652^{+0.040}_{-0.040}$ & - & - & 80.785 & - & - \\
	\midrule	
		\multirow{4}{*}{\vspace{-4pt}CPL}
		& ANN & 140 GRBs & $0.340^{+0.170}_{-0.110}$ & - & $-0.82^{+0.78}_{-0.33}$ & $-0.99^{+0.90}_{-0.58}$ & 53.158 & 3.901 & 9.784 \\
		& GaPP & 140 GRBs & $0.300^{+0.150}_{-0.110}$ & - & $-0.88^{+0.82}_{-0.33}$ & $-1.01^{+0.52}_{-0.91}$ & 40.407 & 3.995 & 9.879 \\
		& ANN & 140 GRBs + 32 OHD & $0.317^{+0.082}_{-0.072}$ & $0.652^{+0.053}_{-0.072}$ & $-0.99^{+0.71}_{-0.45}$ & $-0.96^{+0.58}_{-0.58}$ & 78.953 & 3.622 & 9.917 \\
		& GaPP & 140 GRBs + 32 OHD & $0.314^{+0.086}_{-0.068}$ & $0.652^{+0.051}_{-0.072}$ & $-0.99^{+0.72}_{-0.45}$ & $-0.95^{+0.57}_{-0.57}$ & 81.227 & 3.554 & 9.848 \\
	\bottomrule
	\end{tabularx}
\end{adjustwidth}
\end{table}

\vspace{-6pt}

We also compare models using the Akaike Information Criterion (AIC) and Bayesian Information Criterion (BIC). 
The values of $\Delta\text{AIC}$ and $\Delta\text{BIC}$ relative to the reference model (the $\Lambda$CDM model) are given by: $	\Delta\text{AIC}=\Delta\chi_{\rm min}^2+2\Delta k,~	\Delta\text{BIC}=\Delta\chi_{\rm min}^2+\Delta k\ln N.$	We find that the results of $\Delta \rm AIC$ and $\Delta \rm BIC$  indicate that the $\Lambda$CDM model is favoured respect to the CPL model.				
\section{Conclusions 
}

In this paper, we use an ANN+BNN to calibrate the Amati relation from the Pantheon+ sample to obtain the GRB Hubble diagram with the A219 sample in a cosmology$-$independent way.	Using the ANN method with GRBs at $0.8<z<8.2$ in the A219 sample and 32 OHD,  
we find that the results for the CPL model favor a possible DE evolution ($w_a\neq0$) at the 1 $-$ $\sigma$ confidence region. 
Compared to GP, which imposes strict Gaussian assumptions, ANN eliminates distributional constraints, enabling robust analysis of non$-$Gaussian observational datasets. Our results with GRBs at $1.4\le z\le8.2$ are consistent with previous analyses in \cite{Liang2022,Liu2022b,LZL2023} using GP methods.


We find that the calibration results of the slope in the Amati relation provided by Reichart method are close to the typical value ($b=2$). The physical interpretation of the Amati relation can be provided by a model for the spectral formation of GRB prompt emission \citep{Titarchuk2012}. If the timescale of the GRB prompt emission and its shape for any burst is more or less  the same, then $E_p$ proportional to $E_{\rm iso}^{1/2}$ is seen precisely. 
Frontera et al. \cite{Frontera2012}~have concluded that the Yonetoku relation (the spectral peak energy $E_p$ to the isotropic bolometric luminosity $L_{\rm iso}$) is intrinsic to the emission process, and their results strongly support the reality of both the Amati and Yonetoku relations derived using time-averaged spectra. 

Recently,Ref.\cite{WL2024} presented a sample of long GRBs from fifteen years of the Fermi$-$GBM catalogue with identified redshift, in which the GOLD sample contains 123 long GRBs at $z\le5.6$ and the FULL sample contains 151 long GRBs with redshifts at $z\le8.2$.
In~\cite{Cao2025}, the authors analyzed 151 Fermi$-$observed long GRBs to simultaneously constrain the Amati correlation and cosmological parameters within six spatially flat and non$-$flat dark energy models.
We expect that GRBs could be used to set tighter constraints on cosmological models by using the ANN approach with samples from recent Fermi data \citep{WL2024}.



\newpage
\vspace{6pt}

\authorcontributions{Conceptualization, Z.H. and N.L.; methodology, B.Z.; software, X.L.; validation, Z.H., L.X. and B.Z.; formal analysis, B.Z.; data curation, Z.H., X.L. and B.Z.; writing---original draft preparation, Z.H. , B.Z. and N.L.; writing---review and editing, J.F., Y.L., P.W. and N.L.;  supervision, N.L.;  project administration, N.L.;  funding acquisition,  J.F., Y.L., P.W. and N.L. All authors have read and agreed to the published version of the manuscript.
	
	 
}

\funding{This project was supported by the Guizhou Provincial Science and Technology Foundations (QKHJC$-$ZK[2021] Key 020, QKHJC$-$ZK[2024] General 443 and QKHPT ZSYS [2025] 004). Y. Liu was supported by the NSFC under Grant No. 12373063. P. Wu was supported by the National Natural Science Foundation of China (Grants  No.~12275080) and the Innovative Research Group of Hunan Province (Grant No.~2024JJ1006). J. Feng was supported by the Guizhou Provincial Science and Technology Foundations QKHJC$-$ZK[2022] General 311.}

\dataavailability{Data are contained within the article. 
 } 

\acknowledgments{We thank the anonymous referees for their helpful comments and constructive suggestions.}
\conflictsofinterest{The authors declare no conflicts of interest.} 

\begin{adjustwidth}{-\extralength}{0cm}

\printendnotes[custom]

\reftitle{References}

\PublishersNote{}
\end{adjustwidth}
\end{document}